 \documentclass[final,3p]{elsarticle}
\usepackage{graphicx,subcaption,url}
\usepackage{float}
\usepackage{amsmath,amsthm,amssymb}
\biboptions{sort&compress} 
\newcounter{bla}

\renewcommand{\vec}[1]{\mathbf{#1}}
\newcommand{\be}{\begin{equation}}
\newcommand{\ee}{\end{equation}}
\newcommand{\ba}{\begin{eqnarray}}
\newcommand{\ea}{\end{eqnarray}}

\def\lsim{\raise0.3ex\hbox{$\;<$\kern-0.75em\raise-1.1ex\hbox{$\sim\;$}}}
\def\gsim{\raise0.3ex\hbox{$\;>$\kern-0.75em\raise-1.1ex\hbox{$\sim\;$}}}

\def\theta{\vartheta}

\renewcommand{\vec}[1]{\boldsymbol{#1}}

\newcommand{\apj}{{Astrophys.\ J. }}
\newcommand{\apjl}{{Astrophys.\ J.\ Lett. }}

\journal{Computer Physics Communications}

\begin{document}

\begin{frontmatter}



\title{{\tt ELMAG\,3.01}: A three-dimensional Monte Carlo simulation of
    electromagnetic cascades on the extragalactic background light and in
 magnetic fields}


\author[a]{M.~Blytt}
\author[a]{M.~Kachelrie\ss}
\author[b,c]{and S.~Ostapchenko}

%\cortext[author] {Corresponding author.\\\textit{E-mail address:} firstAuthor@somewhere.edu}
\address[a]{Institutt for fysikk, NTNU, Trondheim, Norway}
\address[b]{Frankfurt Institute for Advanced Studies, Frankfurt, Germany}
\address[c]{D.~V.~Skobeltsyn Institute of Nuclear Physics,
 Moscow State University, Russia}

\begin{abstract}
  The version\,3.01 of {\tt ELMAG}, a Monte Carlo program for the simulation of
  electromagnetic cascades initiated by high-energy photons and electrons
  interacting with extragalactic background light (EBL), is presented. Pair
  production and inverse Compton scattering on EBL photons as well as
  synchrotron losses are implemented using weighted sampling of the cascade
  development. New features include, among others, the implementation of
  turbulent extragalactic magnetic fields and the calculation of
  three-dimensional electron and positron trajectories, 
  solving the Lorentz force equation. As final result of the three-dimensional
  simulations, the program provides two-dimensional
  source images as function of the energy and the time delay of secondary
  cascade particles.
\end{abstract}


\end{frontmatter}

% \linenumbers
%
% Computer program descriptions should contain the following
% PROGRAM SUMMARY.
%
{\bf PROGRAM SUMMARY}\\
  %Delete as appropriate.
%
\begin{small}
\noindent
{\em Manuscript Title:}
{\tt ELMAG\,3.01}: A three-dimensional Monte Carlo simulation of electromagnetic cascades on the extragalactic background light and in turbulent magnetic fields\\
{\em Program Title:} {\tt ELMAG\,3.01} 
%Simulating electromagnetic cascades in the 
%extragalactic  magnetic field and background light.
\\
{\em Journal Reference:}                                      \\
  %Leave blank, supplied by Elsevier.
{\em Catalogue identifier:}                                   \\
  %Leave blank, supplied by Elsevier.
{\em Licensing provisions:}                                   
CC by NC 3.0. % enter "none" if CPC non-profit use license is sufficient.
\\
{\em Programming language:}  Fortran 90                      \\
{\em Computer:}                                               
Any computer with Fortran 90 compiler                  \\ 
{\em Operating system:}  Any system with Fortran 90 compiler                \\
  %Operating system(s) for which program has been designed.
{\em RAM:} 200\,Mbytes, depending on the chosen resolution of the sky images                                              \\
  %RAM in bytes required to execute program with typical data.
{\em Number of processors used:} arbitrary using the MPI version  \\
  %If more than one processor.
{\em Supplementary material:}                                
see \url{http://elmag.sourceforge.net/}
 \\
  % Fill in if necessary, otherwise leave out.
  % Please give some freely chosen keywords that we can use in a
  % cumulative keyword index.
{\em Classification:}        
11.3  Cascade and Shower Simulation, 11.4 Quantum Electrodynamics
                                 \\
  %Classify using CPC Program Library Subject Index, see (
  % http://cpc.cs.qub.ac.uk/subjectIndex/SUBJECT_index.html)
  %e.g. 4.4 Feynman diagrams, 5 Computer Algebra.
{\em Nature of problem:}
Calculation of secondaries produced by electromagnetic cascades on the extragalactic 
background light (EBL), including deflections in magnetic fields.
  %Describe the nature of the problem here.
   \\
{\em Solution method:}
Monte Carlo simulation of pair production and inverse Compton scattering
on EBL photons; weighted sampling of the cascading secondaries; recording of
energy, observation angle and time delay of secondary particles at the
present epoch in a 1.5-dimensional approximation or of sky-maps in a 
three-dimensional approach.
\\
  %Describe the method solution here.
{\em Restrictions:}
Using the three-dimensional approach, the energies of cascade photons should be
above 1\,GeV for $B_{\rm rms} \lsim 10^{-10}$\,G.\\
  %Describe any restrictions on the complexity of the problem here.
%{\em Unusual features:}\\
  %Describe any unusual features of the program/problem here.
%{\em Additional comments:}\\
  %Provide any additional comments here.
{\em Running time:}
20 seconds for $10^4$ photons injected at redshift $z=0.2$ with energy $E=100$\,TeV using one  Intel(R) Core(TM) i7 CPU with 2.8\,GHz in the 1.5-dimensional approximation;  1000 seconds in the three-dimensional approach.
  %Give an indication of the typical running time here.
%* Items marked with an asterisk are only required for new versions
%of programs previously published in the CPC Program Library.\\
\end{small}

%%%%%%%%%%%%%%%%%%%%%%%%%%%%%%%%%%%%%%%%%%%%%%%%%%%%%%%%%%%%%%%%%%%%%%%%%%%%%%
\section{Introduction}

In Ref.~\cite{Kachelriess:2011bi}, the Monte Carlo program  {\tt ELMAG} for
the simulation of
electromagnetic cascades initiated by high-energy photons and electrons
interacting with extragalactic background light (EBL) was introduced. Main
purpose of this program has been the calculation of contributions
to the diffuse extragalactic gamma-ray background and the study of TeV blazars
and the influence of the extragalactic magnetic field (EGMF) on their spectra.
The initial version~1.01 presented in Ref.~\cite{Kachelriess:2011bi} contained
only two EBL models. The spatial structure of turbulent magnetic fields was
approximated by patches of a randomly oriented homogeneous field; the size of
these patches was chosen to correspond to the desired coherence length of
the magnetic field. Deflections and time-delays of charged particles
were then calculated in the small-angle approximation, using a random walk
picture to account for the varying field orientations in different field
patches. As a result of these simplifications, the in general asymmetric
two-dimensional source images were approximated by radially symmetric ones.
In subsequent versions, mainly minor improvements in the code as well as
additional EBL backgrounds were added. However, for potential applications
of the program to studies of gamma-ray sources, using the high-quality data
of future instruments as, e.g., the Cherenkov Telescope Array, a description
of the electromagnetic cascades in the intergalactic space as realistic as
possible is highly desirable.

The version~3.01 of  {\tt ELMAG} presented here includes as an option
to go beyond the limitations
of the 1.5-dimensional simulation, and to employ a more realistic description of
the  turbulent magnetic field. In this approach, the actual trajectories
of electrons\footnote{We call from now on electrons and positrons collectively 
electrons.} are calculated solving the Lorentz force equation using a numerical
solver with adaptive step-size. The turbulent magnetic field is modeled as
an isotropic, divergence-free Gaussian random field with a prescribed
power-law spectrum. Alternatively, the user can read-in his own magnetic
field defined on a chosen grid.
With the change to the version~3.01, also the input of user data has been
updated.  We describe the new format, as well as the output of the
three-dimensional approach, which consists of two-dimensional sky images
visualizing the expected brightness profile of the considered source.
These sky images are functions of the energy and the time delay of the
secondary cascade particles, and depend moreover on the offset angle of the
source with respect to the line-of-sight and the angular profile
of the jet.

%%%%%%%%%%%%%%%%%%%%%%%%%%%%%%%%%%%%%%%%%%%%%%%%%%%%%%%%%%%%%%%%%%%%%%%%%%%%%%
\section{Modeling of three-dimensional cascades}

The simulation of pair production and inverse Compton scattering processes on
EBL photons, of synchrotron losses as well as the  weighted sampling procedure
are essentially unchanged compared to version~1.01, for a description of these
procedures see Ref.~\cite{Kachelriess:2011bi}. In the following, we describe
only the changes
and the additions included for the calculation of three-dimensional cascades.

%%%%%%%%%%%%%%%%%%%%%%%%%%%%%%%%%%%%%%%%%%%%%%%%%%%%%%%%%%%%%%%%%%%%%%%%%%%%%%
\subsection{Turbulent magnetic field}\label{ssec:bturb}

The calculation of the turbulent magnetic field is based on the algorithm
described in Refs.~\cite{1994ApJ...430L.137G,1999ApJ...520..204G}. In that
approach, the magnetic field $\vec{B}(\vec{r})$ at the point $\vec r$ is
obtained as a superposition of transverse Fourier modes with left- or
right-circular polarization,
\be
\vec{B}(\vec{r}) = \sum_{j=1}^{n_k} B_j
\left[ \cos{\alpha_{j}}\vec{\hat{e}}_{x^{'}} +
       {\rm i} h_j\sin{\alpha_{j}}\vec{\hat{e}}_{y^{'}} \right] \,
   {\rm e}^{{\rm i}(k_j \vec{\hat{e}}_{z^{'}} + \beta_j)}.
\label{eq:fourierModes}
\ee
Here, $\vec{\hat{e}}_{i}$ and $\vec{\hat{e}}_{i'}$ denote two sets of Cartesian
unit vector, which are connected by a rotation, $\vec{r^{\prime}}=R \vec{r}$ with
\be
  R(\theta_j, \phi_j)=
  \left[ {\begin{array}{ccc}
   \cos{\theta_j}\cos{\phi_j} & \cos{\theta_j}\sin{\phi_j} & -\sin{\theta_j} \\
   -\sin{\phi_j} & \cos{\phi_j} & 0 \\
   \sin{\theta_j}\cos{\phi_j} & \sin{\theta_j}\sin{\phi_j} & \cos{\theta_j}
  \end{array} } \right] .
\ee
For each mode, two random phases $\alpha_{j}$ and $\beta_{j}$, two rotation
angles $\phi_j$ and $\theta_j$, and the polarization $h_j$ have to be chosen:
For an isotropic magnetic field, $\phi_j$ is uniformly distributed  between
$[0:2\pi]$, while $\cos\theta_j$ is  uniformly distributed between  $[-1:1]$.
The random phases
$\alpha_{j}$ and $\beta_{j}$ are uniformly distributed between $[0:2\pi]$.
The expectation value $\langle h_j\rangle$ of $h_j=\pm 1$ determines the
helicity of the magnetic field, with $|\langle h_j\rangle|=1$ for a
fully left- or right-helical field and $\langle h_j\rangle=0$ for a
field with vanishing helicity. Finally,
the amplitude  $B_j$ of the mode $j$, for a power-law spectrum between
$k_{\min}$ and $k_{\max}$ with root mean square value $B_{\rm rms}$ and slope
$\gamma$, is given by
\be
B_j =  B_{\min}  (k_j/k_{\min})^{-\gamma/2} .
\ee
Here, $ B_{\min}$ denotes the field strength of the lowest Fourier mode, which
is determined by normalizing the total strength of the turbulent field
to $B_{\rm rms}$.

Note also that the coherence length $L_c$ of a turbulent magnetic field with
a power-law like spectrum is connected to its slope and its maximal scale as
\be
 L_{\rm c} = \frac{L_{\max}}{2} \: \frac{\gamma - 1}{\gamma} \:
            \frac{1-(L_{\min}/L_{\max})^{\gamma}}{1-(L_{\min}/L_{\max})^{\gamma-1}}
\simeq
\frac{L_{\max}}{2} \: \frac{\gamma - 1}{\gamma} ,
    \label{eq:correlationlength}
\ee
where the approximation is valid for $L_{\max}=2\pi/k_{\min}\gg L_{\min}=2\pi/k_{\max}$ and $\gamma>1$. For a Kolmogorov spectrum with $\gamma = 5/3$, it follows
$L_{\rm c} = L_{\max}/5$, while  a steep ($\gamma\gg 1$) or monochromatic
($L_{\min}=L_{\max}$) spectrum leads to $L_{\rm c} = L_{\max}/2$.
Electrons will propagate typically in the ballistic regime for the energies of
interest. In this case, $n_k=200$ field modes distributed over two decades
below $L_{\max}$ should be sufficient to ensure an isotropic field with the
desired average helicity. Since for deflections of electrons fluctations
with $k\sim 1/R_L$ are most effective, users have to check if the chosen
value of $k_{\min}$ is sufficiently low for the photon energies simulated.

%%%%%%%%%%%%%%%%%%%%%%%%%%%%%%%%%%%%%%%%%%%%%%%%%%%%%%%%%%%%%%%%%%%%%%%%%%%%%%
\subsection{Calculation of trajectories}

Fluctuations in the EGMF are typically spread over a wide range of scales,
$k_{\min}\ll k_{\max}$. Solving the Lorentz force equation efficiently requires
therefore a scheme with an adaptive step-size solver. The solver employed
is based on routines from Numerical Recipes~\cite{Press1996}, using the
Runge-Kutta formulas from Fehlberg with the Cash-Karp parameters. 
The initial step-size {\tt h1} is set to ${\tt h1} = 10^{-3}{\tt L\_coh}$,
where ${\tt L\_coh}$ denotes the chosen value of the coherence length.

To check the reliability of the numerical scheme, the routine was tested
against the formula for a charged particle being deflected in a turbulent
magnetic field. An electron with energy $E$ propagating the distance $D$
in a turbulent magnetic field with the coherence length $L_c$ is deflected
by the angle~\cite{MiraldaEscude:1996kf}
\begin{equation}
  \Theta \simeq 0.025^{\circ}\sqrt{\frac{D}{L_c}}\frac{L_c}{10\,{\rm Mpc}}\frac{B_{\rm rms}}{10^{-11}\,{\rm G}}\frac{E}{10^{20}\,{\rm eV}}
   \simeq 27.3^{\circ}\frac{\sqrt{DL_c}}{R_L},
    \label{eq:def1}
\end{equation}
where we introduced the Larmor radius $R_L=E/(eB)$ in the last step.

\begin{figure}[H]
\centering
    \begin{subfigure}{0.49\textwidth}
        \centering
        \includegraphics[width=\columnwidth]{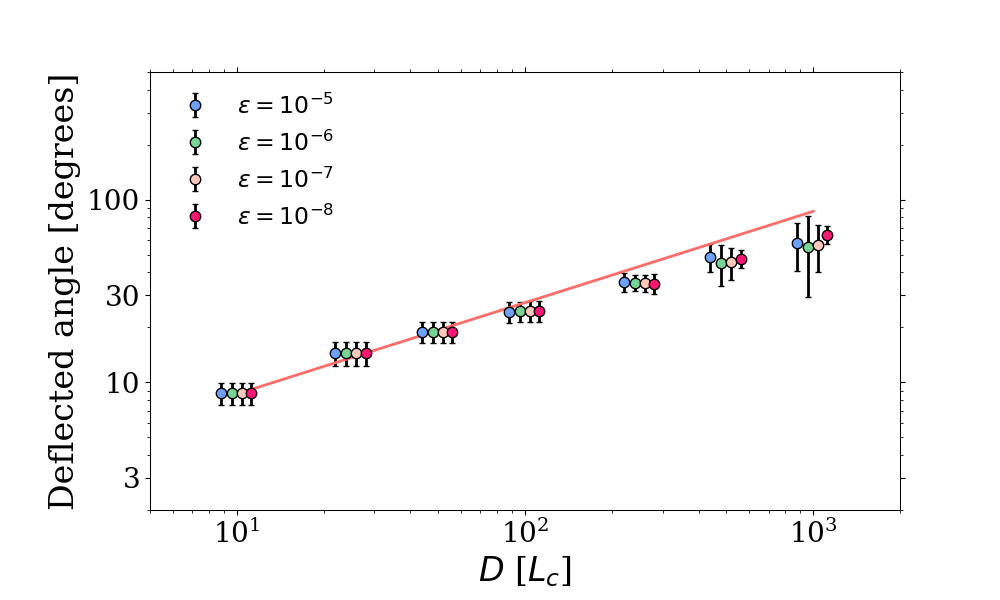}
    \end{subfigure}
    \begin{subfigure}{0.49\textwidth}
        \centering
        \includegraphics[width=\columnwidth]{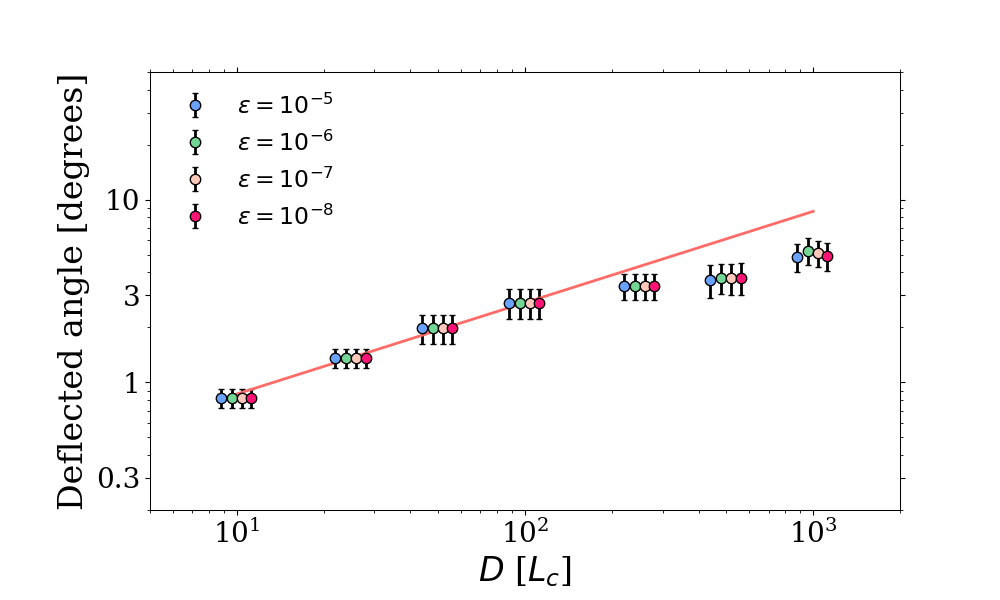}
    \end{subfigure}
    \caption{Deflection angle of electrons as function of the propagation
      distance $D$ in units of $L_c$ for  $R_L=10L_c$ (left panel) and
      $R_L=100L_c$ (right panel) for various values of  {\tt eps}.
      The red line corresponds to the prediction
      of  Eq.~(\ref{eq:def1}).  \label{fig:angleDef} }
\end{figure}

Figure~\ref{fig:angleDef} shows\footnote{For a description of additional tests see also Ref.~\cite{Blytt19}.} the deflection angle obtained with
{\tt ELMAG\,3.01} as function of the
propagation distance $D/L_c$ for two values of the ratio $R_L/L_c$.
Additionally, the error control parameter {\tt eps} has been varied between
$10^{-5}$ and $10^{-8}$; the results for different choices of {\tt eps} were
spread out on the $x$-axis for better readability.
For each case, 100~realisations of the turbulent magnetic field were used
to calculate the mean and the variance of the deflection angle.
The results are in good agreement with Eq.~(\ref{eq:def1})
up to distances $D\sim 100L_c$. Since the interaction length of electrons
is of the order of tens of kpc, i.e.\ smaller than the typical coherence
lengths assumed for the EGMF, we conclude that the numerical precision is
sufficient for our purposes.

As an additional test, we have  compared our results from the three-dimensional
simulation to those from Elyiv {\it et al.}~\cite{Elyiv:2009bx}. For the
comparison, we use a source at the distance 120\,Mpc and a primary photon
energy distribution
\be
E_\gamma^{2}dN_{\gamma}/dE_{\gamma} \propto
\exp(-E_{\gamma}/E_{\max}),
    \label{eq:enDistrEl}
\ee
with an exponential cutoff at $E_{\max}=300$\,TeV as an example. The main
difference between the two simulations is the definition of the magnetic
field. Elyiv  {\it et al.\/} divided the magnetic field into cubic
cells of length 1\,Mpc, where  each cell contains a uniform  field
with random orientation. In contrast, we use a turbulent magnetic field
with a Kolmogorov power-spectrum\footnote{Note that deflections in the magnetic field do not depend on the slope $\gamma$, as long as $\gamma>0$~\cite{Caprini:2015gga}.}
and a coherence length of 1\,Mpc.

\begin{figure}[H]
    \centering
    \includegraphics[width=0.65\columnwidth]{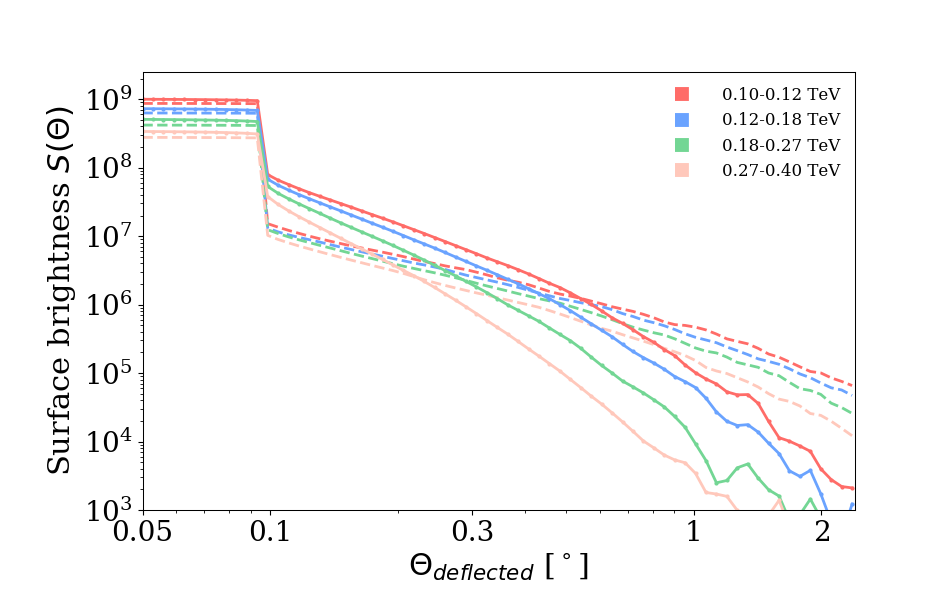}
    \caption{The surface brightness $S(\Theta)$ as function of the deflection angle $\Theta$ for four different energy ranges; dashed lines for a turbulent magnetic field with $B_{\rm rms}=10^{-14}$\,G, while the solid lines are for $B_{\rm rms}=10^{-15}$\,G.}
    \label{fig:angElyiv}
\end{figure}

Figure~\ref{fig:angElyiv} shows the surface brightness $S(\Theta)$ of this
source as a function of the deflection angle $\Theta$ for two magnetic field
strengths. The observed energies of 100.000~photons are summed in four
bins, and the normalization is fixed to the same values as the one in Fig.~6
from Ref.~\cite{Elyiv:2009bx}  for better comparison. 
The plateau visible between
$0.05^{\circ}$ and $0.1^{\circ}$ is caused by ``direct'' photons, which are
smeared out using a point-spread function fitted to data from Fermi-LAT.
The results for the surface brightness $S(\Theta)$ are in good agreement with
those of Ref.~\cite{Elyiv:2009bx}: In particular,
the drops in surface brightness from the plateau to the following bin
coincide well. Our results show, however, a slightly more gradual decline than
the one presented by Elyiv {\it et al.}, which might be connected to the
differences in the magnetic field models employed in the two simulations.

%%%%%%%%%%%%%%%%%%%%%%%%%%%%%%%%%%%%%%%%%%%%%%%%%%%%%%%%%%%%%%%%%%%%%%%%%%%%%%
\subsection{Geometrical setup  of source and observer}\label{ssec:angDelay}

The TeV  emission from active galactic nuclei (AGN) is usually assumed to
be relativistically beamed into a narrow cone with an opening angle
$\Theta_{\rm jet} \simeq 1/(2\Gamma) \simeq 5^\circ [\Gamma/10]$, where
$\Gamma$ is the bulk Lorentz factor of the $\gamma$-ray emitting plasma.
Blazars are a special type of $\gamma$-ray emitting AGN for which the angle
$\Theta_{\rm obs}$ between the line-of-sight and the jet axis satisfies
$\Theta_{\rm obs}\lsim \Theta_{\rm jet}$. Since most observed TeV blazars should
be off-axis with $\Theta_{\rm obs}$ close to $\Theta_{\rm jet}$, the TeV sky-maps
of such sources should be asymmetric.

\begin{figure}[H]
    \centering
    \includegraphics[width=0.85\columnwidth,trim={2cm 20cm 2cm 1cm},clip]{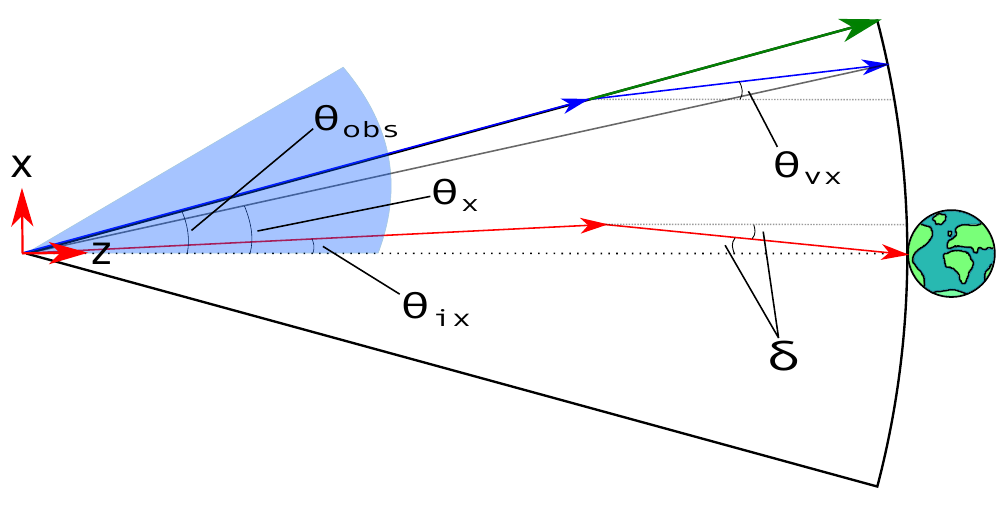}
    \caption{Sketch of our geometrical setup: The blazar jet shown in blue is
      offset by $\Theta_{\rm obs}=\Theta_{\rm jet}$ with respect to the dotted
      line-of-sight. The green arrow represents the initial starting velocity which is chosen to be in the center of the jet. The blue arrows represent the trajectory of the initial particle, the red arrows the rotated path which hits the Earth. }
    \label{fig:geometry}
\end{figure}

In version~3.01 of {\tt ELMAG}, additionally to the opening angle
$\Theta_{\rm jet}$ of the blazar jet, its  offset $\Theta_{\rm obs}$ can be
chosen.  The position of the observer is set at the origin in the $xy$-plane
and at a distance {\tt rcmb} from the source in $z$-direction, independent of
$\Theta_{\rm obs}$. Thus $\Theta_{\rm obs}$ defines the initial velocity vector
of each particle, cf.\ with Fig.~\ref{fig:geometry}. The particle will be
traced until it reaches a distance\footnote{If not otherwise specified, all distances are comoving distances.} $>0.999{\tt rcmb}$ from the origin, i.e.\
until $\sqrt{x^2+y^2+z^2} > 0.999{\tt rcmb}$. After that, it will be projected
onto the sphere of radius {\tt rcmb}. If $f$ denotes the fraction that 
each final velocity component $(v_x, v_y, v_z)$ has to be multiplied 
in order to reach the sphere with radius {\tt rcmb}, then $f$ satisfies
\be
    {\tt rcmb}^{2} = (x+v_x f)^{2} + (y+v_y f)^{2} + (z+v_z f)^{2}.
    \label{eq:projSphere}
\ee
Solving Eq.~(\ref{eq:projSphere}) for $f$ and choosing the smaller absolute
value of the two solutions, gives the corrected final position on the sphere,
$(x_f, y_f, z_f) = (x+v_xf, y+v_yf, z+v_zf)$. If
\be
\Theta_{\rm jet} \leq \arctan\left(\frac{\sqrt{x_f^2+y_f^2}}{z_f}\right)
\label{eq:angleReq}
\ee
is satisfied, the particle trajectory is rotated to hit the Earth;
otherwise the particle is discarded.

Figure~\ref{fig:geometry} visualizes how the rotation in the two-dimensional 
$xz$-plane is performed. The green arrow represents a particle trajectory 
initialized at $\Theta_{\rm obs}=\Theta_{\rm jet}$ with no deflection. The blue 
arrows show a deflected trajectory, with the red arrows representing 
the deflected track rotated to hit the Earth. 
The observation angle $\delta$ can be obtained from the angles 
$\Theta_{vx}=\arctan(v_x/v_z)$ and $\Theta_{x}=\arctan(x/z)$ by 
simple geometrical considerations. 
In order to include a weighted particle distribution inside the jet,
the dependence of $\delta$ on $\Theta_{ix}$ has to be known: Mirroring
$\Theta_x$ at the axis  defined by $\Theta_{\rm obs}/2$ results in 
$\Theta_{ix}=\Theta_{\rm obs}-\Theta_{x}$. Observing then that the triangle
made by  the origin and the two red arrows is the same as the triangle
made by the  origin and the two blue arrows, one can calculate $\delta$
by knowing $\Theta_{\rm obs}$, $\Theta_{vx}$ and $\Theta_{ix}$, viz
$\delta=\Theta_{vx}+(\Theta_{ix}-\Theta_{\rm obs})$.

%%%%%%%%%%%%%%%%%%%%%%%%%%%%%%%%%%%%%%%%%%%%%%%%%%%%%%%%%%%%%%%%%%%%%%%%%%%%%%
\subsection{Stacking and weighting}

The stack holds all secondary particles that have yet to be calculated in
{\tt event}. In version 3.01, the definition of {\tt event} was changed to
\begin{verbatim}
 type one_event
     double precision :: y(6),en,z,w,s,dt
     integer :: icq
  end type one_event 
  type (one_event) act,event(n_max)
\end{verbatim}
Using the three-dimensional option, the variable {\tt  y(6)} contains
the position and velocity vector of each particle, while {\tt s} sums
the distance traveled. Using the 1.5-dimensional option, the variables
{\tt x,the1,the2,xxc,xx} are mapped into {\tt  y(1:5)}.
Calculating the final time delay of each particle can be done by knowing the
total traveled distance when the particle reaches the sphere of radius
{\tt rcmb}. The time delay is then ${\tt dt}= {\tt (s - rcmb)/cy}$, where
{\tt cy} is the speed of light. However, {\tt dt} must be included in the stack to account for the kinematic time delay when the particle is massive. 

The initial {\tt weight} of each particle is now a product of its weight
{\tt w\_e0} in the energy spectrum, its weight {\tt w\_z} in the redshift
distribution, and its weight {\tt w\_jet} in the jet profile.

%%%%%%%%%%%%%%%%%%%%%%%%%%%%%%%%%%%%%%%%%%%%%%%%%%%%%%%%%%%%%%%%%%%%%%%%%%%%%%
\section{Program structure}
The program is distributed among the files {\tt modules301.f90},
{\tt bturb301.f90}, {\tt user301.f90}, {\tt init301.f90}, {\tt elmag301.f90},
and {\tt aux301.f90}. In addition, the two main routines, {\tt main\_sp301.f90}
and {\tt main\_mpi301.f90}, execute a single processor or a MPI simulation,
respectively.
The physics contained in the routines of the new file {\tt bturb301.f90} for
the generation of the turbulent magnetic field has already been
described in subsection~\ref{ssec:bturb}. Additionally, the
file contains the subroutines  {\tt  init\_Bgrid} and
{\tt B\_interpolation(x,Omega)} which read in an EGMF model on a
user-specified Cartesain grid and interpolate these values, respectively.

The changed or added
subroutines and functions in the file {\tt elmag301.f90} which constitute
the core of the program are:
\begin{itemize}
\item 
{\tt subroutine cascade(icq,e00,weight0,z\_in)}\\
Follows the evolution of the cascade initiated by a photon (${\tt icq}=0$) 
or an electron/positron (${\tt icq}=\pm 1$) injected at redshift {\tt z\_in}
with energy {\tt e00} and weight {\tt weight0} until all secondary particles
have energies below the energy threshold {\tt ethr} or reached the sphere of
radius {\tt rcmb}. Depending on the logical variable {\tt three\_dim}, the
routine will calculate three-dimensional trajectories or use the
1.5-dimensional approximation. It will also discard particles 
whose deflection angle exceeds some chosen value.
\item
{\tt subroutine propagate(y,x,e0)}\\
Calculates the trajectory of a charged particle in the turbulent magnetic
field by calling {\tt odeint(y, nvar, 0, x, eps, h1, hmin, nok, nbad, derivs, rkqs, tf)} located in the file {\tt aux301.f90}. The routine chooses a
suitable initial step-size {\tt h1}, depending on the coherence length of the
magnetic field, and the acceptable truncation error {\tt eps} used in the
numerical solver for the trajectory. The particle is tracked for the
interaction length {\tt x},
with {\tt y} containing the particle's position and velocity ($[\,x, y, z, v_x, v_y, v_z]\,$). {\tt hmin} is the minimal step-size for the solver to take and
{\tt nvar}$~=6$ is the number of elements in {\tt y}. 
The outputs {\tt nok} and {\tt nbad} are the number of good and bad 
(but retried and fixed) steps taken.
The inputs {\tt derivs()} and {\tt rkqs()} are external routines that will be
called from inside {\tt odeint()}. The output {\tt tf} is  a logical variable
set to {\tt .false.}, if the solver was not able to solve the ODE given the
required truncation error threshold {\tt eps}. If this happens, the particle
will be discarded. 
\item
{\tt subroutine normalizer(y)}\\
Normalizes the velocity components of {\tt y} so that the absolute value is one.
\item
{\tt subroutine derivs(x,y,dydx)}\\
Is the external subroutine called from {\tt odeint()} and {\tt rkck()}, as a
part of the numerical solver for the Lorentz force equation.  The routine
takes as inputs the
position and velocity of the particle in the six-vector {\tt y}, and outputs
its derivatives {\tt dydx}.  
\item
{\tt subroutine angle\_delay\_3d(y,e0,s,thetax,thetay,dt,w2)}\\
Determines the time-delay {\tt dt} and the observation angle in the
two-dimensional plane from the perspective of a detector on Earth, as
described in subsection \ref{ssec:angDelay}. The outputs {\tt thetax} and
{\tt thetay} are the $\delta$-value from Fig.~\ref{fig:geometry} for the
observed angles in respectively the $xz$-plane and the $yz$-plane. The output
variable {\tt w\_jet} is the weighting variable accounting  for a non-uniform jet
distribution. The latter is calculated by calling the subroutine
{\tt jet\_distribution(the\_s,w\_jet)}, 
where the desired jet distribution
profile is specified.
\end{itemize}

%%%%%%%%%%%%%%%%%%%%%%%%%%%%%%%%%%%%%%%%%%%%%%%%%%%%%%%%%%%%%%%%%%%%%%%%%%%%%%
\section{Example input and output}

The file {\tt user301.f90} is an example file for the input/output subroutines
which should be developed by the user for the desired task. We discuss now the
file contained in the distribution as an example of how the results of the
three-dimensional simulations can be visualized.

%%%%%%%%%%%%%%%%%%%%%%%%%%%%%%%%%%%%%%%%%%%%%%%%%%%%%%%%%%%%%%%%%%%%%%%%%%%%%%
\subsection{Example input}

In previous versions, input parameters were defined as parameters in the module
{\tt user\_variables} or were specified in routines like
{\tt initial\_particle(e0,weight)} and {\tt bemf(r)}. In order to facilitate
the scanning of a grid of parameters, most input parameters are now read
from the files {\tt input\_b}, {\tt input\_src} and {\tt input\_oth}.
The files contain comments which should make their use straight-forward.
For instance, the file {\tt input\_b} starts as
\begin{verbatim}
 2         ! b_model = 1 uniform, 2 turbulent, 3 read from file
 1.d-15    ! B_rms in Gauss
 1d0       ! correlation length L_c/Mpc 
 ...
\end{verbatim}
If the option {\tt b\_model = 3} is chosen, a user-specified model for the
EGMF defined on a Cartesian grid with ${\tt nx\times ny\times nz}$ points
is read in from the file {\tt Bgrid}. The $z$ direction of the grid should
agree with the direction to the considered source.

Additionally, some options can be chosen in the following routines 
of the file {\tt user301.f90}:
\begin{itemize}
\item 
{\tt subroutine  initial\_particle(e0,weight,z)}\\
Assigns the initial photon/electron energy: fixed  {\tt e0}, sampled from
a broken power-law or a user-specified function depending on the value
of the parameter {\tt en\_dist}. Chooses the initial redshift of a particle,
either as fixed {\tt z=z\_max} or sampled from
a distribution {\tt (1+z)**m} up to {\tt z\_max},  depending on the value
of the parameter {\tt z\_dist}.
\item 
{\tt subroutine jet\_distribution(the\_s,w\_jet)}\\
Additional weighting according to a jet distribution, by default a radial
Gaussian distribution with weight one at the center and variance {\tt th\_jet**2}.
\item 
{\tt subroutine psf\_spread(e0,thex,they,weight,dt)}\\
Distributes the detected particle directions according to a PSF,
by default a step function with width equal to {\tt theta\_reg(e0)}.
\item 
{\tt subroutine user\_output(n\_max,n\_proc)}\\
Creates (additionally to the output from previous versions 1.1--2.3) in the
directory
{\tt Output/AngRes} the file
{\tt angle\_matrix} with the  output for the two-dimensional sky-maps.
This directory contains also a python routine for the generation
of plots.
\end{itemize}

%%%%%%%%%%%%%%%%%%%%%%%%%%%%%%%%%%%%%%%%%%%%%%%%%%%%%%%%%%%%%%%%%%%%%%%%%%%%%%
\subsection{Example output and results}

After a particle leaves the cascade, it is stored through the subroutine
{\tt register(e0,thex,they,weight,dt,icq)}. The three-dimensional array
storing the brightness profile from the blazar is
{\tt anobs(n\_bint, n\_binx, n\_biny)}, and depends on the parameters
{\tt n\_bint}, {\tt n\_binx}, {\tt n\_biny}, {\tt n\_bind}, and {\tt shiftx}.
These parameters are respectively the number of time bins for the time delay,
the number of bins in $x$- and in $y$-direction, number of bins per degree, and
how many degrees the grid is shifted in x direction. For a brightness profile
split into four time bins, with a resolution of 30 bins per degree, ranging in
the intervals $x \in[-6^{\circ}, 12^{\circ}]$ and $y \in[-6^{\circ}, 6^{\circ}]$,
one can use the parameters {\tt n\_bint=4}, {\tt n\_binx=541},
{\tt n\_biny=361}, {\tt n\_bind=30} and {\tt shiftx=3}. 

To take into account the typical characteristics of a detector we use the
point spread function (PSF) suggested in Ref.~\cite{Dolag:2010ni},
$\vartheta_{95} \simeq 1.68^{\circ}(E/{\rm GeV})^{-0.77} + 0.2^{\circ}\exp(-10\,{\rm GeV}/E)$, where $E$ is the particle energy and $\vartheta_{95}$ is the
angular containment radius of $95\%$. This function is an analytical
approximation to the PSF from Fermi-LAT, while  $\vartheta_{95}=0.11^{\circ}$
typical for the PSF of a Cherenkov Telescope is used above 300\,GeV.
The function {\tt thereg\_en(e0)} in the program returns the value of
$\vartheta_{95}$. 
Each particle is mapped in the array {\tt anobs}, through the subroutine
{\tt psf\_spread(e0,thex,they,weight,dt)}. In this subroutine, the sizes of
the  bins for the time delay in years can be defined, which are by default:
{\tt 1}: $0 < \tau < 10^{5}$, {\tt 2}: $10^{5} \leq \tau < 10^{6}$, {\tt 3}: $10^{6} \leq \tau < 3 \cdot 10^{6}$, {\tt 4}: $3 \cdot 10^{6} \leq \tau < 10^{7}$ and {\tt 5}: $\tau=0$. This subroutine calls {\tt thereg\_en(e0)}, and
distributes the brightness according a Gaussian with variance
$\sigma\simeq\vartheta_{95}/2$ around the point ({\tt thex}, {\tt they}).

\begin{figure}[H]
    \centering
    \includegraphics[width=0.85\columnwidth]{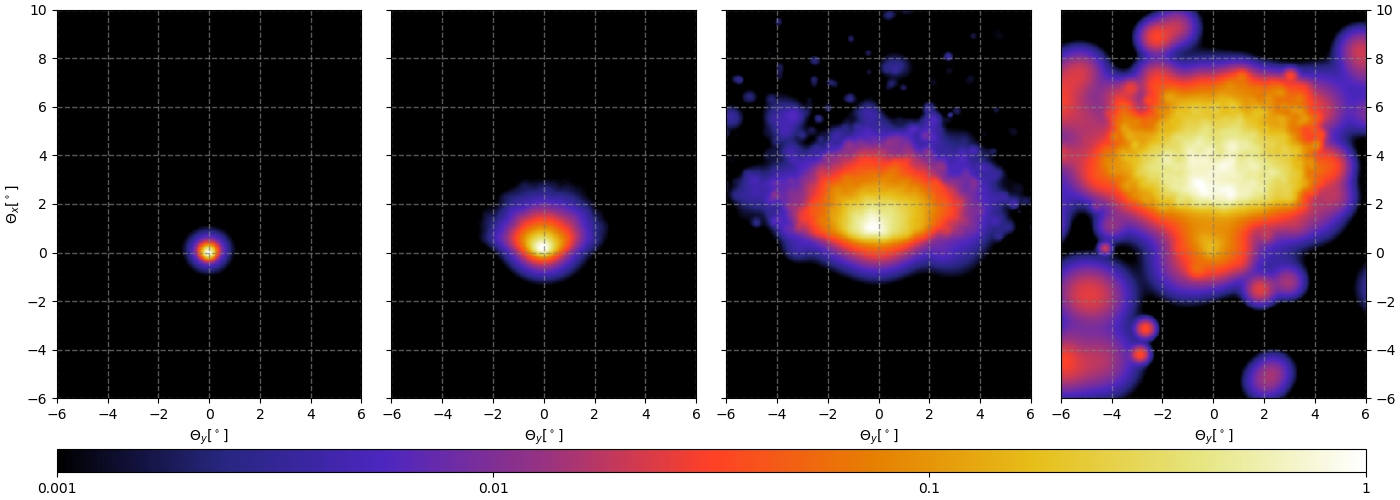}
    \caption{Surface brightness $S$ for photons with energy
      $E_\gamma > 1$\,GeV in the sky region around a blazar emitting a
      photon distribution according to Eq.~(\ref{eq:enDistrEl}); with
      $\Theta_{\rm obs}=\Theta_{\rm jet}=3^{\circ}$, and $B_{\rm rms}=10^{-14}$\,G.
      The four panels correspond to the time bins {\tt 1}, {\tt 2}, {\tt 3}
      and {\tt 4}.  The color code of the surface brightness is shown below.
    \label{fig:surfBright14}}
\end{figure}

\begin{figure}[H]
    \centering
    \includegraphics[width=0.85\columnwidth]{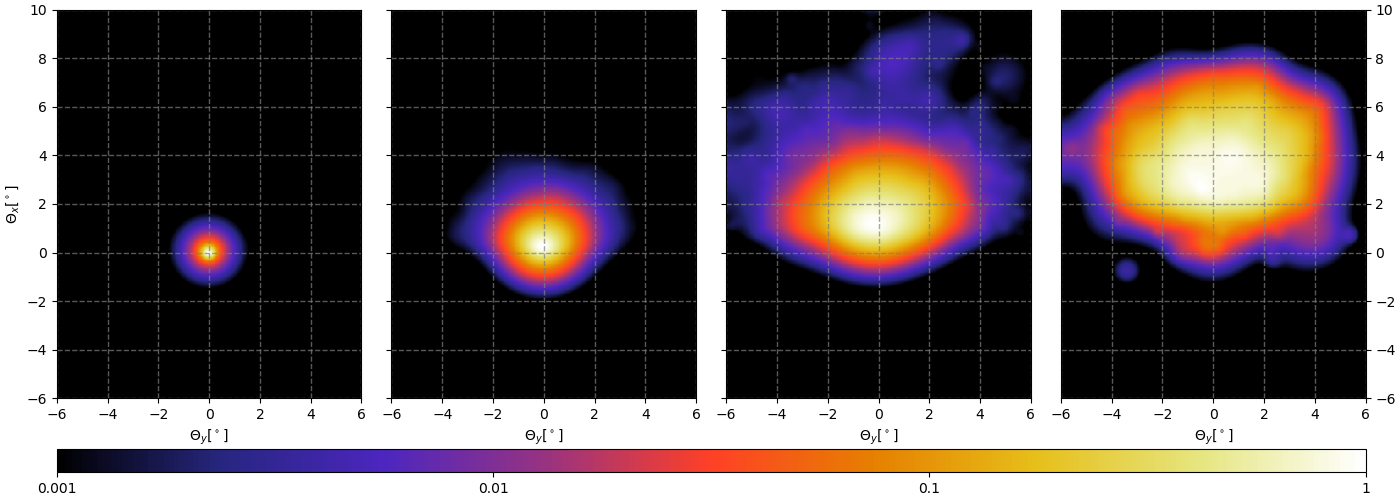}
    \caption{Surface brightness $S$ around a blazar for the
      same parameters as in Fig.~\ref{fig:surfBright14}, but with
      $B_{\rm rms}=10^{-15}$\,G.
    \label{fig:surfBright15}}
\end{figure}

All data arrays exist in two versions, e.g.\ {\tt anobs(n\_bint, n\_binx, n\_biny)} and {\tt anobs\_tot(n\_bint, n\_binx, n\_biny)}. Using MPI\footnote{For information on MPI see e.g.\
``Message Passing Interface Forum. MPI: A Message Passing Interface Standard,
June 1995'' on \url{http://www.mpi-forum.org}.}, 
the former arrays contain the result of a single process, 
which are summed by {\tt call MPI\_REDUCE} into 
{\tt anobs\_tot(n\_bint,n\_binx,n\_biny)},
\begin{verbatim}
  n_array = n_bint*n_binx*n_biny
  call MPI_REDUCE(anobs,anobs_tot,n_array,MPI_DOUBLE_PRECISION,MPI_SUM,0, &
       MPI_COMM_WORLD,ierr)                     ! sum individal arrays spec
\end{verbatim}
Finally, the {\tt subroutine\ user\_output(n\_max,n\_proc)} writes the data
arrays with the results to the files contained in the subdirectory {\tt Output}.
The files {\tt angle\_matrix1} to {\tt angle\_matrix4} contain the normalized surface brightness for the time bins {\tt 1} to {\tt 4}.
Examples for the sky-maps produced with the Python routine contained in
the directory {\tt Output/AngRes} are shown in Figs.~\ref{fig:surfBright14}
and ~\ref{fig:surfBright15}. They show the surface brightness around a
blazar for two different strengths of the turbulent magnetic field,
using one million injected photons.

%%%%%%%%%%%%%%%%%%%%%%%%%%%%%%%%%%%%%%%%%%%%%%%%%%%%%%%%%%%%%%%%%%%%%%%%%%%%%%
\section{Summary}

We have presented the new features in {\tt ELMAG\,3.01} which include the
generation of (helical) turbulent magnetic fields and the three-dimensional
tracking of cascade particles. This allows the user to produce two-dimensional
sky-images of sources with an arbitrary jet offset angle and jet profile. The
format of the new input and output as well as a few tests have been described.
Potential energy losses of electrons caused by plasma instabilities are
included as a new option in {\tt ELMAG\,3.02}, which is described in the
addendum,

%%%%%%%%%%%%%%%%%%%%%%%%%%%%%%%%%%%%%%%%%%%%%%%%%%%%%%%%%%%%%%%%%%%%%%%%%%%%%%
\section*{Acknowledgements}
\noindent
We would like to thank Manuel Meyer and  Andrey Saveliev for contributing
additional EBL backgrounds, and Kumiko Kotera,  Manuel Meyer, Marco Muzio
and Foteini Oikonomou for useful suggestions which improved {\tt ELMAG}.
S.O.\ acknowledges support from the project OS\,481/2-1 of the 
Deutsche Forschungsgemeinschaft.

%%%%%%%%%%%%%%%%%%%%%%%%%%%%%%%%%%%%%%%%%%%%%%%%%%%%%%%%%%%%%%%%%%%%%%%%%%%%%%
%%%%%%%%%%%%%%%%%%%%%%%%%%%%%%%%%%%%%%%%%%%%%%%%%%%%%%%%%%%%%%%%%%%%%%%%%%%%%%

%\bibliographystyle{elsarticle-num}
%\bibliography{elmag}

%%%%%%%%%%%%%%%%%%%%%%%%%%%%%%%%%%%%%%%%%%%%%%%%%%%%%%%%%%%%%%%%%%%%%%%%%%%%%%
%%%%%%%%%%%%%%%%%%%%%%%%%%%%%%%%%%%%%%%%%%%%%%%%%%%%%%%%%%%%%%%%%%%%%%%%%%%%%%
\newpage

\section*{Addendum: Implementation of energy losses due to plasma instabilities in {\tt ELMAG} 3.02}

The authors of Ref.~\cite{Broderick:2011av} first suggested that the beam of
$e^+e^-$ pairs generated by a TeV blazar through pair production is prone to
plasma instabilities. If the energy losses caused by the growth of unstable
plasma modes can compete with those due to inverse Compton scattering, the
standard evolution of an electromagnetic cascade is modified. In particular,
the suppression of the GeV emission from bright TeV blazars, which
has been used to constrain the intergalactic magnetic field,
may be caused alternatively by such plasma instabilities.

Analytical studies of the growth rate of plasma instabilities have been
performed only in the linear regime~\cite{Broderick:2011av,Miniati:2012ge,2012ApJ...758..102S,Schlickeiser:2013eca}.
Moreover, these studies generally employ
additional simplifications as assuming, e.g., an uniform  density of background
electrons. Therefore, they cannot address the question  when and how
the exponential growth of instabilities stops under realistic conditions.
Numerical simulations, on the other hand, cannot be performed in the parameter
range relevant for TeV blazar beams~\cite{Sironi:2013qfa,Vafin:2018kox}.
Therefore, numerical results from PIC
simulations have to be rescaled, relying on the validity of analytical scaling
relations~\cite{Pohl:2019nvw}. As a result, the predictions by different
authors for the growth rate of plasma instabilities vary vastly.

Reference~\cite{AlvesBatista:2019ipr} collected five models for the energy
loss rate due to plasma instabilities and compared their impact on the energy
spectra of TeV blazars. The formulas for the energy losses predicted in these
models, as summarized in~\cite{AlvesBatista:2019ipr}, have been implemented in
{\tt ELMAG}~3.02. The new file {\tt plasma302.f90} contains the loss rates
for the different models. The energy loss due to plasma instabilities is
added to the synchrotron losses in the
function {\tt eloss} contained in {\tt elmag302.f90}. The user chooses in the
file {\tt input\_pls} the model and its parameters
\begin{verbatim}
 N    ! Instability model; N - None, A - Broderick et al. (2012), B - Miniati & 
Elyiv (2013), C - Schlickeiser et al. (2012, 2013), D - Sironi & Giannios (2014), 
E - Vafin et al. (2019) 
 1d4  ! temperature/K of IGM
 1d-7 ! density/cm^3 of IGM 
 1d45 ! total isotropic-equivalent luminosity in erg/s
 1.d0 ! overall scaling factor f_scale
\end{verbatim}
The last parameter allows the user to rescale the energy
loss rates due to plasma instabilities by the overall factor  {\tt f\_scale}. 

\begin{figure}[H]
    \centering
    \includegraphics[width=0.45\columnwidth]{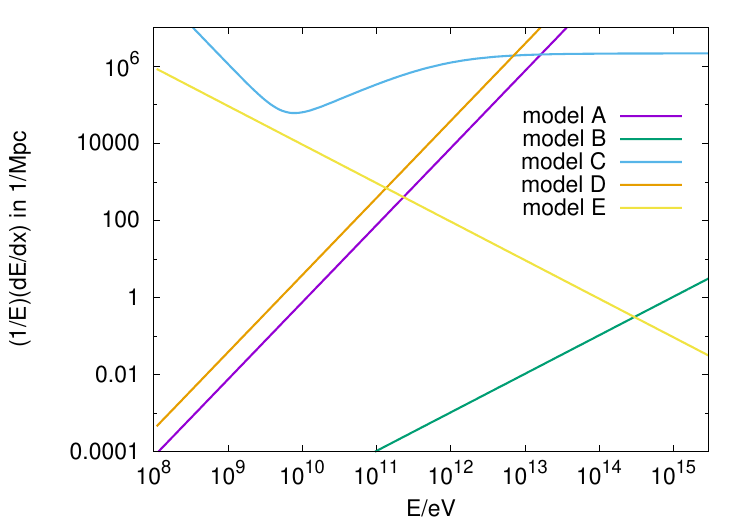}
    \includegraphics[width=0.45\columnwidth]{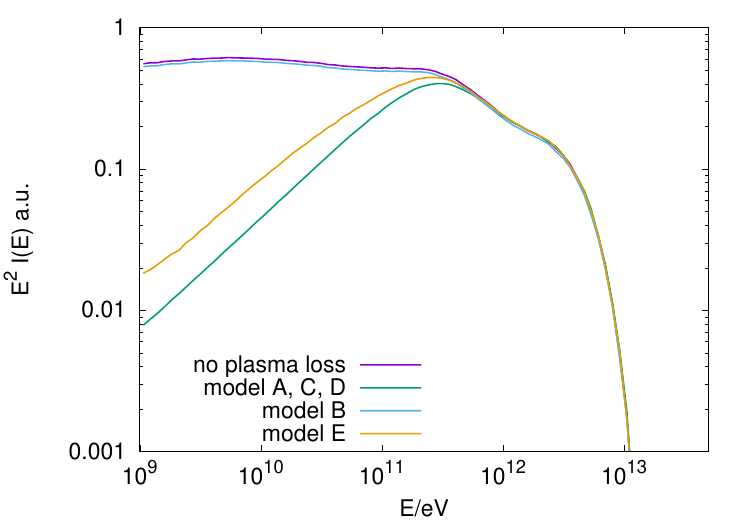}
    \caption{Left: Energy loss rates using model A~\cite{Broderick:2011av},
      B~\cite{Miniati:2012ge},
      C~\cite{2012ApJ...758..102S,Schlickeiser:2013eca},
      D~\cite{Sironi:2013qfa} and
      E~\cite{Vafin:2018kox} for the growth rate of plasma instabilities.
      Right: The resulting photon spectra $E^2I(E)$ as function of energy.
      All for $T_{\rm IGM}=10^4$\,K, $n_{\rm IGM}=10^{-7}/$cm$^3$  and
      isotropic luminosity $L=10^{45}$erg/s. 
    \label{fig:Eloss}}
\end{figure}

In the left panel of Fig.~\ref{fig:Eloss}, we compare the resulting energy
loss rates %to the one of inverse Compton scattering
of the different models\footnote{Note that the
  corresponding Fig.~1 in Ref.~\cite{AlvesBatista:2019ipr} contains
  errors~\cite{PC}.}. In the case of model C, the growth rates in the two
different density regimes, $n_{\rm IGM}\ll n_{\rm cr,C}$ and
$n_{\rm IGM}\gg n_{\rm cr,C}$, are discontinuous at $n_{\rm cr,C}$.
In order to obtain a smooth behavior, we connected the two regimes
using logistic functions.

In the right panel of Fig.~\ref{fig:Eloss}, we show the resulting photon
spectra $E^2I(E)$ for a source at $z=0.14$ with injection spectrum
$dN/dE\propto E^{-1.2}\exp(-E/E_0)$ and $E_0=5$\,TeV. In the case of model~B,
plasma losses do not affect the cascade process and, therefore,
the photon spectrum  in model~B is indistinguishable from the one without
plasma losses. In case of the other models, plasma losses stall
the  electromagnetic cascade and, as result, the slope of the photon spectrum
is similar to the injection spectrum. Note that in contrast
to Ref.~\cite{AlvesBatista:2019ipr}, the spectrum in model C shows no
upturn at low energies, after the discontinuity has been smoothed out.

%%%%%%%%%%%%%%%%%%%%%%%%%%%%%%%%%%%%%%%%%%%%%%%%%%%%%%%%%%%%%%%%%%%%%%%%%%%%%%
\section*{Acknowledgements}
\noindent
We would like to thank Jacob Daniel Benestad for implementing
the energy losses induced by plasma instabilitities into {\tt ELMAG}.

%\bibliographystyle{elsarticle-num}
%\bibliography{elmag}

\end{document}